# Fine Discrimination of Analog Patterns by Nonlinear Dendritic Inhibition

Kenji Morita   and   Kazuyuki Aihara

*Department of Complexity Science & Engineering, University of Tokyo, Tokyo 113-8656, Japan*

Recent experiments revealed that a certain class of inhibitory neurons in the cerebral cortex make synapses not onto cell bodies but at distal parts of dendrites of the target neurons, mediating highly nonlinear dendritic inhibition. We propose a novel form of competitive neural network model that realizes such dendritic inhibition. Contrary to the conventional lateral inhibition in neural networks, our dendritic inhibition models don't always show winner-take-all behaviors; instead, they converge to "I don't know" states when unknown input patterns are presented. We derive reduced two-dimensional dynamics for the network, showing that a drastic shift of the fixed point from a winner-take-all state to an "I don't know" state occurs in accordance with the increase in noise added to the stored patterns. By preventing misrecognition in such a way, dendritic inhibition networks achieve fine pattern discrimination, which could be one of the basic computations by inhibitory connected recurrent neural networks in the brain.

Nonlinearity is one of the most prominent features of neural systems. Especially, nonlinear input summation on dendrites of neurons has been attracting both experimental and theoretical attentions, leading to fruitful findings of what single neurons can do [1]. However, possibly emerging features of neuronal networks composed of such highly nonlinear components have not been thoroughly examined, except for a few interesting studies [2].

Recent experiments using intracellular simultaneous recordings indicate that there exists a wealth of dendritic inhibition in the brain, including the cerebral cortex, mediated by a certain class of inhibitory neurons, non-FS cells [3,4]. While inhibition added to one arbor of the dendrites is considered to cancel excitation on the same arbor, it has little effect on the other arbors. Thus such dendritic inhibition is much more nonlinear than cell body-directed somatic inhibition [1]. We introduce a novel type of competitive neural network model with such nonlinear dendritic inhibition. We examine dynamical properties and computational abilities of the model, to show they are quite different from conventional competitive neural networks that always behave in well-known winner-take-all fashions.

*Models of two types of inhibition.*---Threshold linear networks have been used to describe the dynamics of neural activity in terms of firing rates [6,7]. Among them, networks with global inhibition have been shown to have winner-take-all properties and have been used as a computational architecture of the basic local circuits of the brain [8]. Such networks consist of an input layer, an output layer, excitatory feed-forward connections, and inhibitory recurrent connections in the output layer (see Fig. 1 (a)). They can be described as follows:

$$\frac{dx_j}{dt} = -x_j + \left[\sum_{i=1}^{m} w_{ji} I_i - \beta \sum_{h=1, h \neq j}^{n} v_{jh} x_h \right]^+$$

where $[y]^+ = \max\{y, 0\}$ denotes rectification, $I_i$ the intensity of the $i$-th component of the input vector $\mathbf{I}$, $x_j$ the firing rate of the $j$-th neuron, $w_{ji} (\geq 0)$ the excitation from the $i$-th input to the $j$-th neuron, $v_{jh} (\geq 0)$ the inhibition which the $h$-th neuron gives the $j$-th neuron, $\beta$ a parameter representing the relative strength of inhibition, $m$ the dimension of the input vector $\mathbf{I}$, and $n$ the number of neurons.

In this model, all the excitatory and inhibitory inputs are linearly summed. Therefore, It is considered to be a model of inhibition onto cell bodies (somata), which we refer as the somatic lateral inhibition (SLI) model in this paper. On the other hand, dendritic inhibition should be modeled so that inhibition is added to each dendritic arbor before summation so as to restrict their effects only to the arbor to which they have been added. To describe the property, we introduce a novel type of competitive neural network, namely, the dendritic lateral inhibition (DLI) model (see Fig. 1 (b)), which can be represented as follows:

$$\frac{dx_j}{dt} = -x_j + \sum_{i=1}^{m} \left[ w_{ji} I_i - \beta \sum_{h=1, h \neq j}^{n} w'_{hi} x_h \right]^+$$

where $w'_{hi} \geq 0$ denotes the inhibition that the $h$-th neuron gives to every feed-forward connection from the $i$-th input to any other neurons, which are set to equal to $w_{hi}$, for the sake of simplicity, because they share the same pre- and post- pair in the *reverse* order and so possibly have the equal strengths after appropriate learning based on the correlation of both sides' activities [9].

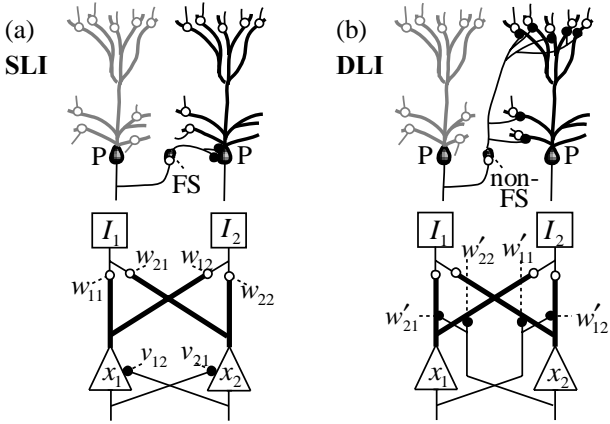

FIG. 1. (a) Somatic lateral inhibition (SLI). (b) Dendritic lateral inhibition (DLI). The case of $m=n=2$ are illustrated. White circles: excitatory synapses, Black circles: inhibitory ones, P: pyramidal cells.

*Comparison of dynamics.*---We will compare the general dynamical properties of the SLI and the DLI models with a random weight matrix $W=(w_{ji})$ for various inputs. Specifically, we take $m$ random values from the uniform distribution in $[0\ 1]$, and normalize them so that they satisfy $\sum_{i=1}^{m} w_{ji}=1$ for $\forall j(=1,\cdots,n)$, as post-synaptic normalization conditions.

SLI models with symmetrical connections have been shown to have Lyapunov-like functions that ensure the existence of attracting equilibrium points [7]. Here we consider the case of uniform inhibition. In this case, if the inhibition strength is strong enough ($\beta>1$), all of the equilibrium points will be the winner-take-all (WTA) states in which only one neuron is active [7]. Among them, the equilibrium point in which the neuron receiving the maximum total input is the only winner has the widest basin including the origin of the coordinates, so that under appropriate conditions, the SLI network works as a maximum input selector; *i.e.*, a neuron that receives the maximum input becomes the unique winner.

First, we assume that an input pattern $\mathbf{I}$ is parallel to a transpose of one (*e.g.* the $k$-th) of the row vectors of the weight matrix $W=(w_{ji})$, $\mathbf{w}_k=(w_{k1}\cdots w_{km})$, which corresponds to excitatory synaptic weights of the $k$-th neuron: $\mathbf{I}=(\sum_{i=1}^{m} I_i)\mathbf{w}_k^t$. Then, the $k$-th neuron receives the total input

$$\mathbf{w}_k \mathbf{I} = (\sum_{i=1}^{m} I_i)\mathbf{w}_k\mathbf{w}_k^t = (\sum_{i=1}^{m} I_i)|\mathbf{w}_k|^2$$

On the other hand, the other neuron (*e.g.* the $j$-th neuron, $j\neq k$) receives the total input

$$\mathbf{w}_j\mathbf{I} = m(\tfrac{1}{m}\sum_{i=1}^{m} w_{ji})(\tfrac{1}{m}\sum_{i=1}^{m} I_i) + m\cdot\mathrm{Cov}(\mathbf{w}_j^t,\mathbf{I})$$
$$= \tfrac{1}{m}\sum_{i=1}^{m} I_i + \mathrm{Cov}(m\mathbf{w}_j^t,\mathbf{I}),$$

where $\mathrm{Cov}(m\mathbf{w}_j,\mathbf{I})=\tfrac{1}{m}\sum_{i=1}^{m}(mw_{ji}-\sum_{h=1}^{m}w_{jh})(I_i-\tfrac{1}{m}\sum_{h=1}^{m}I_h)$ denotes covariance between $m\mathbf{w}_j$ and $\mathbf{I}$. The second equality is derived from $\sum_{i=1}^{m} w_{ji}=1$. Since $W$ is a random matrix, $\mathbf{I}$ is *ideally* uncorrelated with any row vectors of $W$ except $\mathbf{w}_k$, *i.e.*, $\mathrm{Cov}(m\mathbf{w}_j^t,\mathbf{I})$ converges to 0 in probability when $m\to\infty$ by the law of large numbers. This results in that the $k$-th neuron receives maximum input among all other neurons, because when $m$ is large enough

$$\frac{\mathbf{w}_k\mathbf{I}}{\mathbf{w}_j\mathbf{I}} \approx m|\mathbf{w}_k|^2 = m|\mathbf{w}_k|^2\left|(\tfrac{1}{\sqrt{m}}\ \cdots\ \tfrac{1}{\sqrt{m}})^t\right|^2$$
$$> m\left(\mathbf{w}_k\cdot(\tfrac{1}{\sqrt{m}}\ \cdots\ \tfrac{1}{\sqrt{m}})^t\right)^2 = m\cdot\tfrac{1}{m} = 1$$

Consequently, the $k$-th neuron almost always becomes the winner when the input dimension $m$ is large. In other words, the network can *discriminate* the input pattern parallel to $\mathbf{w}_k$. In this sense, we refer $\mathbf{w}_k$ ($k=1,\cdots,n$) as the "stored" patterns of the network.

Next, let us consider which neuron will become a winner if an input $\mathbf{I}$ is not exactly parallel to any of $\mathbf{w}_j^t$, but has positive correlation with one of those. Specifically, we consider an input

$$\mathbf{I}=\frac{\mathbf{w}_k^t+\mu\xi}{\tfrac{1}{m}\sum_{i=1}^{m}(w_{ki}+\mu\xi_i)}$$

where $\xi=(\xi_1\ \cdots\ \xi_m)^t$ represents random noise added to the stored pattern $\mathbf{w}_k^t$ satisfying

$$\sum_{i=1}^{m}\xi_i=1\ \text{and for}\ ^\forall j,\ \mathrm{Cov}(\mathbf{w}_j^t,\xi)\to 0\ (m\to\infty)$$

and $\mu$ is the amount of noise. The denominator of (eq. #) represents normalization so that the average of the components of $\mathbf{I}$ is equal to 1. If $\mathbf{w}_j$ and $\xi$ are *ideally* random and independent, the $k$-th neuron always receives the largest total input so that it becomes the winner. However, practically the input dimension $m$ is finite, and so there are correlations between $\mathbf{w}_j$ and $\xi$. Thus, the probability that another neuron will become the winner instead increases as the noise level increases, as shown in Fig. 2. (a), (b). In the extreme case, if the input pattern is random, all of the neurons are equally likely to become the winner because the expectation values of their total input are equal.

In this way, the SLI model can be regarded as a somewhat robust but non-specific pattern discriminator; that is, if a stored pattern with small noise is presented, then its corresponding neuron becomes active, whereas if a severely noise-distorted or an unknown pattern is presented, then a randomly chosen neuron becomes active, leading to misrecognition (type-two error).

The dynamical behaviors of the DLI models, however, are quite different (see Fig.2. (c), (d)). When a stored pattern (parallel to one of $\mathbf{w}_k$) with small noise is presented, only the corresponding neuron becomes highly active in a WTA-*like* manner, similar to the case of the SLI models; but not exactly WTA, because the losers have small nonzero values if noise exists. However, when a severely noise-distorted or an unknown pattern is presented, the DLI model no longer converges to a WTA-*like* state, but to another state, namely an "I don't know" state, in which all the neurons have little activity. In this way, the DLI model can be regarded as a sophisticated pattern discriminator: the neuron with the highest activity indicates which of the stored patterns is nearer to the currently presented one, but its level of activity indicates how good the match is. A comparison of the abilities of the SLI and the DLI model in a pattern discrimination is summarized in Fig. 2. (e), (f).

*Reduced two-dimensional dynamics of the DLI networks.*---In order to explain why the DLI models show such interesting behaviors, let us reduce the whole dynamics of the model to two dimensions. First, we divide all the neurons into two groups, one potential winner (let it be the 1-st neuron for simplicity) and the other potential losers. Then (eq. #) becomes

$$\frac{dx_1}{dt} = -x_1 + \sum_{i=1}^{m}\left[w_{1i}I_i - \beta\sum_{h=2}^{n}w'_{hi}x_h\right]^+ ,$$

$$\frac{dx_j}{dt} = -x_j + \sum_{i=1}^{m}\left[w_{ji}I_i - \beta(w'_{1i}x_1 + \sum_{h=2, h\neq j}^{n}w'_{hi}x_h)\right]^+ .$$

Here, we introduce the "mean activity" of the losers $\bar{x} = \frac{1}{n-1}\sum_{j=2}^{n}x_j$ and replace the losers' true activities in the above equations with $\bar{x}$, so that approximately

$$\begin{cases}\frac{dx_1}{dt} = -x_1 + f_1(\bar{x}) \\ \frac{d\bar{x}}{dt} = -\bar{x} + f_2(x_1, \bar{x})\end{cases},$$

where

$$f_1(\bar{x}) = \sum_{i=1}^{m}\left[w_{1i}I_i - \beta\sum_{h=2}^{n}w'_{hi}\bar{x}\right]^+ ,$$

$$f_2(x_1,\bar{x}) = \sum_{i=1}^{m}\left[w_{ji}I_i - \beta\left(w'_{1i}x_1 + \sum_{h=2, h\neq j}^{n}w'_{hi}\bar{x}\right)\right]^+ .$$

represent the total non-negative input to the potential winner and the losers, respectively [10].

Figure 3. (a) plots numerically calculated $f_1$ and $f_2$ without noise ($\mu=0$), or with heavy noise ($\mu=8$). A significant change can be seen in the form of $f_2$: its foot along the $x_1$-axis shifts upwards as the noise level increases, which means that the potential losers can

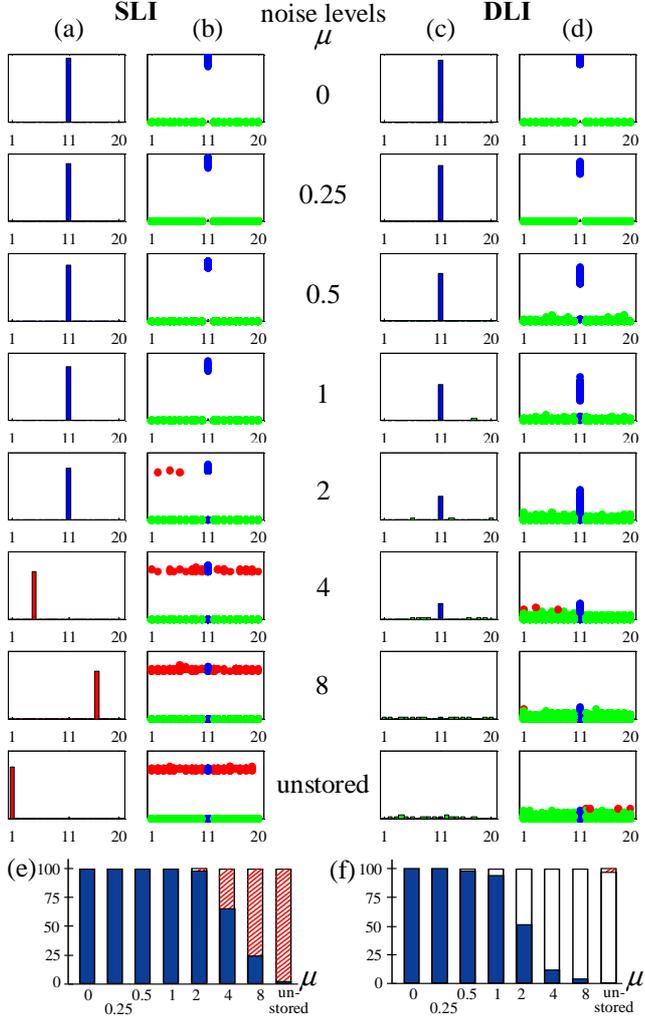

FIG. 2. (a-d) Responses of the SLI or the DLI model to patterns with various noise levels ($\mu = 0 \sim 8$) or to an unstored random pattern. Horizontal axis: index of neurons ($j = 1,\cdots,20$). Vertical axis: steady state firing rates. (a,c) Examples of single trials in which the pattern $\mathbf{w}_{11}^t$ corresponding to *the 11-th neuron* (numbered on the axis) with various amounts ($\mu$) of noise is presented. (b,d) Overpaintings of 100 trials. The parameter values and conditions: $m=100$, $n=20$, and $\beta=5$; $W$ and $\mathbf{I}$ are reset for each trial; initial values of $x_j$ obey $N[0.1,0.01]$. (e,f) Discrimination abilities (vertical axis, %) of the SLI model (e) and the DLI model (f) for various noise levels (horizontal axis). Filled: success ($x_{11} > \text{mean}(x_j) + 5\cdot\text{std}(x_j)$), White: "I don't know" ($x_j < 0.2$ for $\forall j$), Stripe: misrecognition (other cases).

receive positive input even if the potential winner has a high activity. This change in $f_2$ together with the change in $f_1$ results in a large shift in the steady state (the attracting fixed point) of the $x_1$-$\bar{x}$ dynamics (see Fig. 3. (b)); from a state in which only $x_1$ has a high activity (the WTA-like state) to a state where both $x_1$ and $\bar{x}$ have little activities (the "I don't know" state). Fig. 3. (c) shows the shift in the $x_1$ value of the steady state with noise. Note that the values match very well with the average of the simulations, indicating that this reduced two-dimensional dynamics well explain the dynamical behavior of the original model.

*Discussions and Conclusions.*---There are several evidences that there exist plenty of dendritic inhibition in the brain, which may act in distinct circuitries from somatic one [3-5]. Moreover, these two inhibition modes are found to be differentially regulated by some neural modulators like acetylcholine and dopamine [4]. Although it remains to be seen how these two inhibitory systems differentiate and/or cooperate in their functions, such a fine discrimination of input patterns achieved by the DLI model as described above could be the basic function of dendritic inhibition distinct from winner-take-all competition mediated by somatic one. For example, it may be used for humans or animals to represent objects with critical meanings for them, such as faces or body odors. This possibility, as well as the others, for example that dendritc inhibition may work more slowly than somatic one because of the distance from somata and then these two inhibition modes work in different coding schemes (see [11]), should be examined in future studies. Aside from it, since the DLI model is described using simple piecewise-linear dynamics that is easily implemented by analog VLSI [12], overall results here are sensible and potentially useful in terms of a biologically inspired computational device, for example, a pattern discriminator based on biometrics.

In conclusion, we have developed a novel mathematical model of neural circuits, which is motivated by recent observations about richness of highly nonlinear dendritic inhibitions. We have shown that the model doesn't always converge to a winner-take-all state but instead converge to an "I don't know" state for severely noise-distorted or unknown patterns, achieving highly specific pattern discrimination. In addition, we have shown that a mean-fields like approach dividing all the neurons into either a potential winner or losers is an effective way of analyzing dynamical behaviors of such competitive neural networks.

This study is partially supported by a Grant-in-Aid No. 15016023 for Scientific Research on Priority Areas (2) Advanced Brain Science Project and Superrobust Computation Project in COE Program on Information Science and Technology Strategic Core from the Ministry of Education, Culture, Sports, Science, and Technology, Japanese Government.

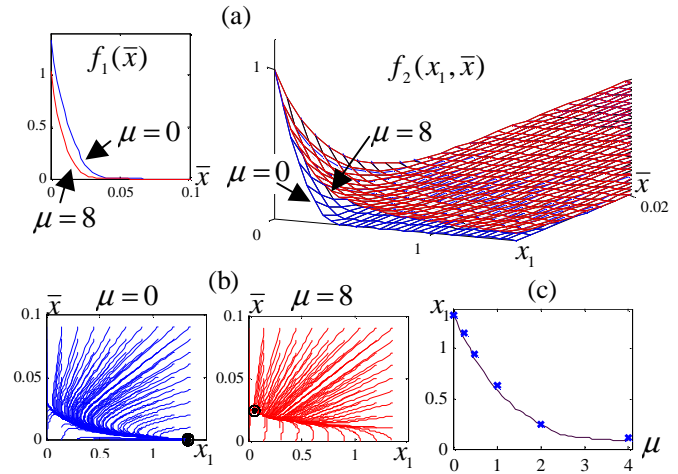

FIG. 3. (a) Forms of $f_1$ (left) and $f_2$ (right). $f_1$ shifts downwards, whereas $f_2$ shifts upwards as noise $\mu$ increases from $0$ to $8$. (b) Trajectories of reduced two-dimensional dynamics of the DLI model (left: $\mu=0$, right: $\mu=8$). Black circles indicate the steady states. (c) Change in firing rate of the winner at the steady states (the vertical axis) plotted against noise level $\mu$ (the horizontal axis). The theoretical values (line) are comparable to the average values of the simulations (crosses).